\documentclass[twocolumn,showpacs,prl,letterpaper,amsmath,amssymb,superscriptaddress]{revtex4}

\usepackage{graphicx}
\usepackage{dcolumn}
\usepackage{bm}
\usepackage{color}
\definecolor{red}{rgb}{0.8,0.0,0.0} 

\begin{document}

\preprint{PRL/123-QED}

\title{Solvable model for chimera states of coupled oscillators}

\author{Daniel M. Abrams}
\affiliation{Department of Earth, Atmospheric, and Planetary Sciences, 54-621, Massachusetts Institute of Technology, Cambridge, MA 02139, USA.}

\author{Renato E. Mirollo}
\affiliation{Department of Mathematics, Boston College, Chestnut Hill, MA 02467, USA.}

\author{Steven H. Strogatz}
\email{strogatz@cornell.edu}
\affiliation{Department of Theoretical and Applied Mechanics, Cornell University, Ithaca, NY 14853, USA.}

\author{Daniel A. Wiley}
\affiliation{Department of Mathematics, University of Maryland, College Park, MD 20742, USA.}

\date{\today}

\begin{abstract}
Networks of identical, symmetrically coupled oscillators can spontaneously split into synchronized and desynchronized sub-populations.  Such chimera states were discovered in 2002, but are not well understood   theoretically.  Here we obtain the first exact results about the stability, dynamics, and bifurcations of chimera states by analyzing a minimal model consisting of two interacting populations of oscillators.   Along with a completely synchronous state, the system displays stable chimeras, breathing chimeras, and saddle-node,  
Hopf and homoclinic bifurcations of chimeras. 
\end{abstract}



\maketitle

Many creatures sleep with only half their brain at a time~\cite{rattenborg}.  
Such unihemispheric sleep was first 
reported in dolphins and other sea mammals, and has now 
been seen in birds and inferred in lizards~\cite{mathews}.  When brain waves are 
recorded, the awake side of the brain shows desynchronized
electrical activity, corresponding to millions of neurons oscillating out of 
phase, whereas the sleeping side is highly synchronized. 

From a physicist's perspective, unihemispheric sleep suggests 
the following (admittedly, extremely idealized) problem: 
What's the simplest system of two oscillator populations, 
loosely analogous to the two hemispheres, such that one synchronizes while the other does not? 

Our work in this direction was motivated by a series 
of recent findings in nonlinear dynamics~\cite{kuramoto2002, abrams, shima, omelchenko, sethia, PREs}. 
In 2002, Kuramoto and Battogtokh reported that arrays of   
nonlocally coupled oscillators could spontaneously split into 
synchronized and desynchronized sub-populations~\cite{kuramoto2002}.  
The  existence of such ``chimera states'' 
came as a surprise, given that the  oscillators were identical 
and symmetrically coupled. 
On a one-dimensional ring~\cite{kuramoto2002, abrams} the chimera took the form of 
synchronized domain next to a desynchronized one.  
In two dimensions, it appeared as a strange new kind of spiral wave~\cite{shima}, with phase-locked
oscillators in its arms coexisting with phase-randomized oscillators in its core---a circumstance made possible only by the nonlocality of the coupling. 
These phenomena were unprecedented in studies of pattern formation~\cite{hoyle} and synchronization~\cite{syncrefs} in physics, chemistry, and biology, and remain poorly understood.

Previous mathematical studies of chimera states have 
assumed that they are statistically stationary~\cite{kuramoto2002,shima, abrams, omelchenko, sethia}.  
What has been lacking is an analysis of their dynamics, stability, and bifurcations. 
 
In this Letter we obtain the
first such results by considering the simplest model that supports
chimera states: a pair of oscillator populations in which each oscillator is 
coupled equally to all the others in its group, and less 
strongly to those in the other group.   For this model
we solve for the stationary chimeras and delineate where
they exist in parameter space.  An unexpected finding is that chimeras
need not be stationary.  They can breathe.  Then the phase 
coherence in the desynchronized population waxes and wanes, while  
the phase difference between the two populations begins to wobble.

The governing equations for the model are
\begin{equation}\label{eq:gov}
  \frac{d \theta^\sigma_i}{dt} = \omega +  
    \sum^2_{\sigma^{\prime} = 1} \frac{K_{\sigma \sigma^{\prime }}}{N_{\sigma^{\prime}}}
    \sum^{N_{\sigma^{\prime}}}_{j=1} \sin ( \theta^{\sigma^{\prime}}_j 
    - \theta^\sigma_i - \alpha )
\end{equation}
where $ \sigma = 1, 2$ and $ N_\sigma$ is the number of oscillators in population $
\sigma $. The oscillators are assumed identical, so the frequency $\omega$ and phase lag $\alpha$ are the same for all of them. The strength of the coupling from oscillators in $ \sigma^\prime$ onto those in $ \sigma $ is given by $K _{\sigma \sigma^{\prime }}$. To facilitate comparison with earlier work, we suppose that $ K _{11} = K_{22} = \mu > 0 $, and $ K _{12} = K
_{21} = \nu > 0 $, with $ \mu > \nu $. Thus, the coupling within a
group is stronger than the coupling between groups. This corresponds to the assumption~\cite{kuramoto2002, abrams, shima} of a nonlocal coupling that decreases with distance. By rescaling time, we may set $\mu + \nu=1$. It also proves useful to define the parameters $A=\mu-\nu$ and $\beta=\pi/2-\alpha$, because, as we'll show, chimeras exist only if these quantities are small enough.  

\begin{figure}[t] 
   \includegraphics[width=3in, clip, angle=0]{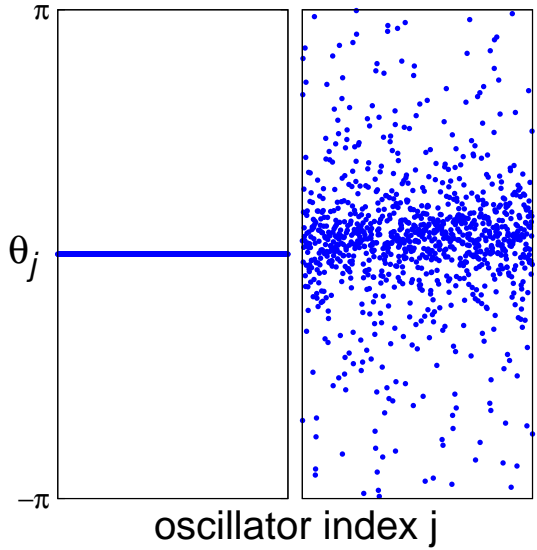} \\
   \caption{Snapshot of a chimera state, obtained by numerical integration of \eqref{eq:gov} with $\beta=0.1$,  $A=0.2$, and $N_1=N_2=1024$. The two populations are shown side by side.}
   \label{fig:chimera}
\end{figure}

Simulations of Eq.~\eqref{eq:gov} display two types of behavior.  For many initial conditions, the system approaches the synchronized state where all $\theta$'s are equal. Otherwise it evolves to a chimera state (Fig.~\ref{fig:chimera}).  The oscillators in group~1 are in sync; those in group~2 are not. 

\begin{figure}[t] 
   \includegraphics[width=3in, clip, angle=0]{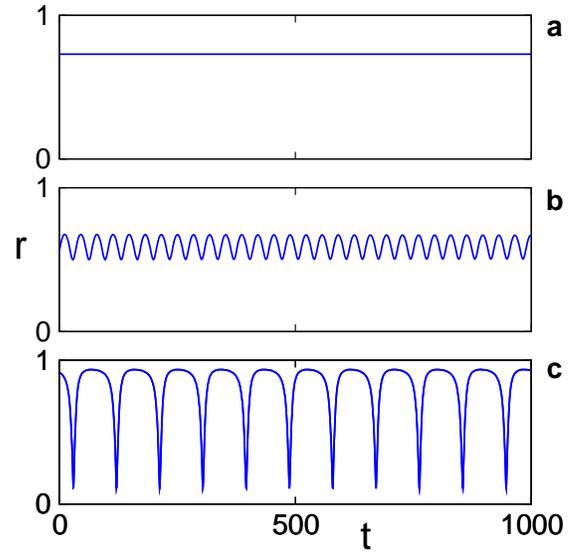} \\
   \caption{Order parameter $r$ versus time. In all three panels, $N_1=N_2=128$ and $\beta=0.1$. (a) $A=0.2$: stable chimera; (b) $A=0.28$: breathing chimera; (c) $A=0.35$: long-period breather. Numerical integration began from an initial condition close to the chimera state, and plots shown begin after allowing a transient time of 2000 units.}
   \label{fig:rt}
\end{figure}

Figure~\ref{fig:rt} illustrates the dynamics of chimera states.  We plot the phase coherence of the desynchronized population, as quantified by the order parameter 
$r(t)= \vert\langle e^{i \theta_j(t)} \rangle_2 \vert$,  
where the angle brackets denote an average over all oscillators in population $\sigma =2$. In Fig.~\ref{fig:rt}(a) the order parameter remains constant, except for slight fluctuations due to finite-size effects.  Thus, this chimera is stable and statistically stationary.  However, if we increase $ \mu$ (the coupling within a population) relative to $\nu$ (the coupling between populations), the stationary state can lose stability.  Now the order parameter pulsates, and the chimera starts to breathe (Fig.~\ref{fig:rt}(b)).  The breathing cycle lengthens as we increase the disparity $A=\mu-\nu$ between the couplings (Fig.~\ref{fig:rt}(c)).  At a critical disparity, the breathing period becomes infinite.  Beyond that, the chimera disappears and the synchronized state becomes a global attractor.

To explain these results, we analyze Eq.~\eqref{eq:gov} in the continuum limit where $ N_\sigma \rightarrow \infty $ for $\sigma = 1, 2$.  Then Eq.~\eqref{eq:gov} gives rise to the continuity equations
\begin{equation}\label{eq:continuity}
\frac{\partial f^\sigma }{\partial t} + \frac{\partial }{\partial
\theta } ( f ^\sigma v ^\sigma ) = 0 ,
\end{equation}
where $ f^\sigma (\theta, t) $ is the probability density of
oscillators in population $ \sigma$, and $ v^\sigma (\theta, t) $ is their
velocity, given by
\begin{equation}\label{eq:velocity}
v ^\sigma ( \theta    , t) =   \omega + \sum^2 _{\sigma ^{\prime} =
1} K _{\sigma \sigma ^{\prime }} \int \sin  ( \theta ^{\prime}   -
\theta    - \alpha ) f ^{\sigma ^{\prime}} ( \theta ^{\prime} , t) \,
d \theta  ^{\prime}.
\end{equation}
(Note that we dropped the superscripts on $ \theta$ to ease the notation. Thus, $ \theta $ means $ \theta ^\sigma $ and $ \theta^\prime $ means
$ \theta ^{\sigma ^{\prime }}$.)
If we define a complex order parameter
\begin{equation}\label{eq:complexorder}
z_\sigma (t) = \sum^2_{\sigma^{\prime} = 1}    K _{\sigma \sigma
^{\prime }} \int      e ^{i \theta ^{\prime}}  f^{\sigma ^{\prime}}
( \theta ^\prime, t) \, d \theta ^{\prime},
\end{equation}
then $ v^\sigma  (\theta) $ simplifies to
\begin{eqnarray}\label{  6}
 v ^\sigma (\theta, t) & = & \omega + \textrm{Im} [ e ^{-i \theta} e ^{- i
\alpha} z _\sigma (t) ]\nonumber\\
& = & \omega + \frac{1}{2i} (z _\sigma  e ^{-i \alpha} e ^{-i \theta
} - z ^\ast _\sigma e ^{i \alpha} e ^{i \theta}),
\end{eqnarray}
where the $ \ast $ denotes complex conjugate.

Following Ott and Antonsen~\cite{ott}, we now consider a special class
of density functions $ f^\sigma $ that have the form of a Poisson
kernel. The remarkable fact that Ott and Antonsen discovered is that
such kernels satisfy the governing equations {\em exactly}, if a certain
low-dimensional system of ODEs is satisfied. In other words, for this
family of densities, the dynamics reduce from infinite dimensional to
finite (and low) dimensional. (Numerical evidence suggests that all attractors 
lie in this family, but proving this remains an open problem.) 

Specifically, let
\begin{equation}\label{eq:poisson}
f ^\sigma (\theta, t) = \frac{1}{2 \pi} \left\{ 1+ \left[\sum^\infty
_{n=1} ( a _\sigma (t) e ^{i \theta} )^n + c.c. \right] \right\}.
\end{equation}
What's special here is that we use the same function $ a_\sigma
(t)$ in all the Fourier harmonics, except that $ a_\sigma $ is raised
to the $ n^{\rm th}$ power in the $ n^{\rm th}$ harmonic.
Inserting this $f^\sigma $ into the governing equations, one finds
that this is an exact solution, as long as
\begin{equation}\label{eq:amplitude}
\dot{a}_\sigma + i \omega a _\sigma + \frac{1}{2} \left[ a ^2 _\sigma
z _\sigma e ^{-i \alpha} - z ^\ast _\sigma  e ^{i \alpha}\right] = 0.
\end{equation}
Instead of infinitely many amplitude equations, we have just one.
(It's the same equation for all $n$.)

To close the system, we express the complex order parameter $
z _\sigma $ in terms of $ a_\sigma $. Inserting the Poisson kernel~\eqref{eq:poisson}
into Eq.~\eqref{eq:complexorder}, and performing the integrals, yields
\begin{equation}\label{eq:z}
z_\sigma (t) = \sum^2_{\sigma^{\prime} = 1}    K _{\sigma \sigma
^{\prime }}  \,     a ^\ast _{\sigma^{\prime}} (t),
\end{equation}
by orthogonality.
Thus the amplitude equations become
%
\begin{eqnarray}\label{eq:amp}
 0 &=& \dot{a}_1 + i \omega a_1 + \frac{1}{2} a^2_1 \left( K_{11} 
 a^\ast_1 + K_{12} a^\ast_2 \right) e ^{- i \alpha}  \nonumber\\
 & & - \frac{1}{2}\left(K_{11} a_1 + K_{12} a_2 \right)e^{i \alpha} 
\end{eqnarray}
and similarly for $ \dot{a}_2$, with 1's and 2's interchanged.

Rewrite the amplitude equations using polar coordinates $ \rho$ and $
\phi$, defined by $ a_\sigma = \rho_\sigma e ^{- i \phi_{\sigma}},\enspace
\sigma = 1, 2 $.  
(The negative sign is included in the definition of
$ \phi $ so that the Poisson kernel $ f^\sigma $ converges to $
\delta (\theta - \phi _\sigma )$, not $ \delta (\theta + \phi _\sigma
)$, as $ \rho \rightarrow 1 $ from below. Thus $ \phi _\sigma $ can
be interpreted as the ``center'' of the density $ f^\sigma $, and $
\rho_\sigma $ measures how sharply peaked it is.) 
Then Eq.~\eqref{eq:amp} becomes
\begin{eqnarray}\label{eq:polarampeqns}
0 & = & \dot{\rho}_1 + \frac{\rho_1^2 - 1}{2} \left[ \mu \rho _1
\cos \alpha + \nu \rho _2 \cos \left( \phi _2 - \phi _1 - \alpha
\right) \right] \nonumber\\
0 & = &  -{\rho} _1\dot{\phi}_1  + \rho_1 \omega \nonumber\\
& & - \frac{1+
\rho^2_1}{2} \left[ \mu \rho _1 \sin \alpha + \nu \rho _2 \sin \left(
\phi _1 - \phi _2 + \alpha\right) \right]
\end{eqnarray}
and similarly for $ \dot{\rho}_2  $ and $ \dot{\phi} _2$, with 1's
and 2's interchanged.

Now consider the case of chimera states, for which one
population is in sync while the
other is not. Taking $\sigma=1$ to be the synchronized population,
we set $ \rho_1 \equiv 1$, corresponding to a $ \delta$-function for
that group. Note that $ \rho_1 \equiv 1 $ satisfies the governing
equations for all time, since $ \dot{\rho} _1 = 0 $ when $ \rho_1 =
1$. Hence the condition $ \rho _1 \equiv 1 $ defines an invariant
manifold, on which the dynamics reduce to
\begin{eqnarray}\label{eq:3dsystem}
 \dot{\phi}_1 & = & \omega - \mu \sin \alpha - \nu r \sin ( \psi +
\alpha ) \nonumber\\
\dot{r} & = & \frac{1-r^2}{2} \left[ \mu r \cos \alpha + \nu \cos
(\psi - \alpha) \right]\nonumber\\
\dot{\phi}_2 & = & \omega - \frac{1 + r^2}{2 r} \left[ \mu r \sin
\alpha + \nu \sin ( \alpha - \psi ) \right]
\end{eqnarray}
where we've defined $ r = \rho_2 $ and $ \psi = \phi _1 - \phi _2$.
The $(r, \psi)$ dynamics decouple, yielding a 2-$D$
system given by
\begin{eqnarray}\label{eq:2dsystem}
\dot{r} & = & \frac{1-r^2}{2} \left[ \mu r \cos \alpha + \nu \cos
(\psi - \alpha) \right]\nonumber \\
\dot{\psi} & = & \frac{1+r^2}{2r} \left[ \mu r \sin \alpha - \nu \sin
(\psi - \alpha) \right]-\mu \sin \alpha \nonumber\\
& & - \nu r \sin ( \psi + \alpha )~.
\end{eqnarray}

\noindent This system has a trivial fixed point $r=1,\psi=0$ (perfectly synchronized state)
which always exists.  
The non-trivial fixed points  
correspond to stationary chimera states, in which the local
order parameters $ \rho _1 (t) \equiv 1$ and $ \rho_2 (t) = r(t) $
remain constant, as does the phase difference $ \psi (t) = \phi _1
(t) - \phi_2(t) $, despite the fact that the individual microscopic
oscillators in population $ \sigma = 2 $ continue to move in a
desynchronized fashion.

\begin{figure}[t] 
    \includegraphics[width=3in, clip, angle=0]{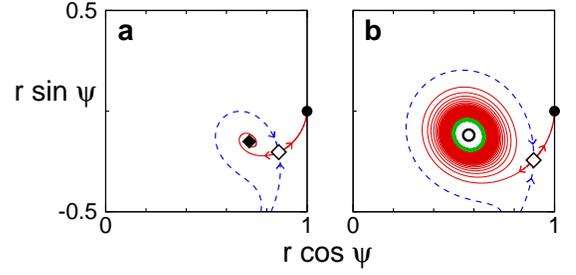}
        \caption{Phase portraits for Eq.~\eqref{eq:2dsystem}, regarding $r$ and $\psi$ as polar coordinates. Parameters as in Figs.~\ref{fig:rt}(a),(b), respectively. (a) Stable chimera (solid diamond). (b) Breathing chimera, shown as a stable limit cycle (thick curve) about unstable chimera (open dot). In both panels: open diamond, saddle chimera; thin solid line, unstable manifold; dashed line, stable manifold;  solid dot, stable synchronized state.} 
    \label{fig:phaseplane}
\end{figure}

Figure~\ref{fig:phaseplane} plots typical phase portraits for~\eqref{eq:2dsystem}.
Figure~\ref{fig:phaseplane}(a) shows a stable chimera state coexisting with the stable synchronized state; the basin boundary between them is defined by the stable manifold of a saddle chimera. As we increase 
the disparity $A$ between the couplings $\mu$ and $\nu$, the chimera becomes less stable and eventually undergoes a supercritical Hopf bifurcation, creating a stable limit cycle (Fig.~\ref{fig:phaseplane}(b)), the counterpart of the breathing chimera of Fig.~2(b).

Additional phase plane analysis (not shown) reveals two other bifurcations.  With further increases in $A$, the limit cycle expands and approaches the saddle.  Meanwhile, its period lengthens, which accounts for the behavior seen earlier in Fig.~\ref{fig:rt}(c).  At sufficiently large $A$ the cycle touches the saddle point and destroys itself in a homoclinic bifurcation. On the other hand, if $A$ is decreased from its value in Fig.~\ref{fig:phaseplane}(a), the stable and saddle chimeras in Fig.~\ref{fig:phaseplane}(a) approach each other, and eventually coalesce and annihilate in a saddle-node bifurcation.  
   
\begin{figure}[t] 
   \includegraphics[width=2.8in, clip, angle=0]
            {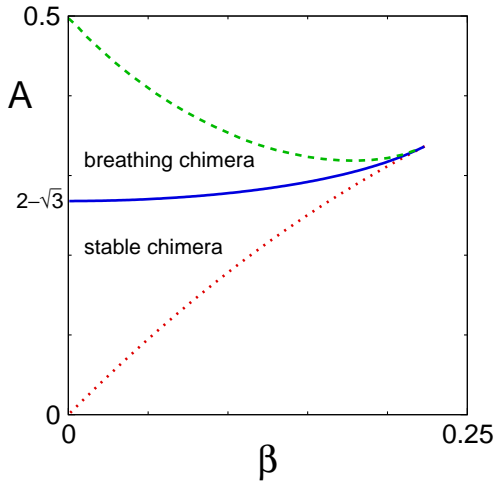} \\
   \caption{Stability diagram for chimera states.  Bifurcation curves: saddle-node (dotted) and supercritical Hopf (solid), both found analytically;  homoclinic (dashed), found numerically. All three curves intersect at a Takens-Bogdanov point $(\beta, A) = \scriptstyle \left(\cos^{-1} \sqrt{\sqrt{13}/8 + 1/2}, \frac{3-\sqrt{2\sqrt{13}-5}}{3+\sqrt{2\sqrt{13}-5}} \right) \displaystyle = (0.2239, 0.3372)$.} 

   \label{fig:stabdiag}
\end{figure}

Figure~\ref{fig:stabdiag} summarizes the bifurcations and stability regions for the system. 
In the rest of the paper we outline the analysis leading to these results. 

To calculate the fixed points for Eq.~\eqref{eq:2dsystem}, 
we set $ \dot{r} = 0 $
and $ r \neq 1 $ (since group $\sigma=2$ is desynchronized) and 
obtain $ \mu r \cos \alpha + \nu \cos ( \alpha - \psi) =
0 $. Substituting $\mu=(1+A)/2,\nu=(1-A)/2$ and $\beta=\pi/2-\alpha$ into 
$ \dot{r} = 0 $ and solving for $A$ yields
\begin{equation}\label{eq:Aparametric}
A = \frac{\sin (\beta + \psi) + r \sin \beta}{\sin(\beta + \psi) -r \sin \beta}
\end{equation}
at a fixed point.
Then imposing $ \dot{\psi} = 0$ and using the expression for $A$ above, we find
\begin{equation}\label{eq:rparametric}
r = \sqrt{\frac{\sin (2 \beta + \psi)}{\sin (2 \beta - \psi) + 2 \sin \psi}} . 
\end{equation}

Equations~\eqref{eq:Aparametric} and \eqref{eq:rparametric} together parametrize all the fixed points.  They 
define a surface, or equivalently, a two-parameter family.  
Sweeping $\beta$ and $\psi$ yields the corresponding $r$ and $A$ values.
The resulting surface has two sheets that join along a fold; its projection 
onto the $(\beta, A)$ plane defines the curve of saddle-node bifurcations.   

To calculate the saddle-node curve, we linearize~\eqref{eq:2dsystem} 
about a fixed point, and set the determinant of the Jacobian to zero.  This implies 
\begin{equation}\label{eq:snpsi}
\sin \beta + \frac{\sin (2 \beta + \psi) [ \sin (\beta - 2 \psi ) + 2
\sin (\beta + 2 \psi)]}{\sin (2 \beta - \psi) + 2 \sin \psi} = 0, 
\end{equation}
where we've used Eqs.~\eqref{eq:Aparametric} and \eqref{eq:rparametric} to
simplify the determinant.  
Solving~\eqref{eq:snpsi} for $ \psi(\beta) $ yields two roots, but one of them implies $ r > 1 $ and
hence is spurious; the correct root is
\begin{equation}
\psi = - 2 \beta - 2 \beta ^2 + 2 \beta ^3 + \frac{11 \beta ^4}{3} -
12 \beta ^5 - \frac{3271 \beta^6}{180} + O(\beta^7).
\end{equation}
This is then substituted into \eqref{eq:rparametric} to yield $r(\beta)
$, which in turn yields $ A (\beta)$, via \eqref{eq:Aparametric}. In
this way we obtain the saddle-node curve
\begin{equation}\label{  21}
A_{SN} (\beta) =  2\beta - 2\beta ^2 -\frac{7 \beta^3}{3}+ \frac{20
\beta^4}{3}  + \frac{181 \beta^5}{60}+ O(\beta ^6), 
\end{equation}
which matches the numerical curve shown in Fig.~\ref{fig:stabdiag}.  

To find the Hopf curve, we set the trace of the
Jacobian to zero, which gives 
$\psi =  -\frac{1}{2} \sin ^{-1} ( 2 \sin 2 \beta).$
Repeating the procedure above leads to an exact parametric equation for the Hopf
curve. Its leading order behavior is
\begin{equation}\label{eq:hopfapprox}
A_{\rm H} (\beta) =  2 - \sqrt{3} + \left(4\sqrt{3} - 6\right)
\beta ^2 + \left( \frac{26}{\sqrt{3}}-10\right)\beta ^4 + O(\beta
^6) .
\end{equation}

Future work should investigate whether breathing chimeras exist for the one- and two-dimensional arrays of oscillators studied previously~\cite{kuramoto2002,shima, abrams, omelchenko, sethia}.  Are the stability diagrams for such systems similar to Fig.~4? Do chimeras also exist if the oscillators are non-identical~\cite{syncrefs,ott,barreto} or arranged in complex networks~\cite{netrefs}?  It would also be worth looking for experimental examples of chimera states.  Candidate systems include arrays of lasers~\cite{laser} and chemical~\cite{chemical} or electrochemical~\cite{electrochemical} oscillators.

We thank E.~Ott and T.~M. Antonsen for sending us their preprint. Research supported by NSF grant DMS-0412757 to S.H.S. and a Mathematical Sciences Postdoctoral Research fellowship to D.M.A.

\end{document}